\documentclass[structabstract]{aa}  

\usepackage{graphicx}
\usepackage{txfonts}
\usepackage{natbib}
\bibpunct{(}{)}{;}{a}{}{,} 
\begin{document}
   \title{Constraining mixing processes in stellar cores using asteroseismology}

   \subtitle{Impact of semiconvection in low-mass stars}

   \author{V. Silva Aguirre,\inst{1}
          \and
          J. Ballot,\inst{2}
	  \and
          A.M. Serenelli,\inst{1,3}
	  \and
          A. Weiss\inst{1}
          }

   \institute{Max Planck Institute for Astrophysics, Karl-Schwarzschild-Str. 1, 85748, Garching bei M\"{u}nchen, Germany \\
              \email{[vsilva;aldos;weiss]@mpa-garching.mpg.de}
         \and
             Laboratoire d'Astrophysique de Toulouse-Tarbes, Universit\'e de Toulouse, CNRS, 14 avenue E. Belin, 31400 Toulouse, France \\
             \email{jballot@ast.obs-mip.fr}
         \and
             Instituto de Ciencias del Espacio (CSIC-IEEC), Facultad de Ci\`encies, Campus UAB, 08193 Bellaterra, Spain
             }

   \date{Received ---; accepted ---}

  \abstract
   {The overall evolution of low-mass stars is heavily influenced by the processes occurring in the stellar interior. In particular, mixing processes in convectively unstable zones and overshooting regions affect the resulting observables and main sequence lifetime.}
   {We aim to study the effects of different convective boundary definitions and mixing prescriptions in convective cores of low-mass stars, and to discriminate the existence, size, and evolutionary stage of the central mixed zone by means of asteroseismology.}
   {We implemented the Ledoux criterion for convection in our stellar evolution code, together with a time-dependent diffusive approach for mixing of elements when semiconvective zones are present. We compared models with masses ranging from 1 M$_{\sun}$ to 2 M$_{\sun}$ computed with two different criteria for convective boundary definition and including different mixing prescriptions within and beyond the formal limits of the convective regions. Using calculations of adiabatic oscillations frequencies for a large set of models, we developed an asteroseismic diagnosis using only $l=0$ and $l=1$ modes based on the ratios of small to large separations $r_{01}$ and $r_{10}$ defined by Roxburgh \& Vorontsov (2003). We analyzed the sensitivity of this seismic tool to the central conditions of the star during the main sequence evolution.}
   {The seismic variables $r_{01}$ and $r_{10}$ are almost linear in the expected observable frequency range, and we show that their slope depends simultaneously on the central hydrogen content, the extent of the convective core, and the amplitude of the sound-speed discontinuity at the core boundary. By considering about 25 modes and an accuracy in the frequency determinations as expected from the CoRoT and \textit{Kepler} missions, the technique we propose allows us to detect the presence of a convective core and to discriminate the different sizes of the homogeneously mixed central region without the need of a strong a priori for the stellar mass.}
   {}

   \keywords{Stars: evolution -- Convection -- Asteroseismology -- Stars: oscillations}

   \titlerunning{Probing mixing in stellar cores using asteroseismology}

   \maketitle
%
\section{Introduction}\label{s:intro}
Understanding the physical processes dominating stellar interiors and correctly characterizing their impact on the evolution of stars is one of the main goals of stellar astrophysics. For a long time it has been acknowledged that different mixing processes in stars such as convection, microscopic diffusion and rotational mixing strongly influence the overall evolution of stars \citep[e.g.][and references therein]{jpz92,amgm00,at94,st98}, with subsequent effects in observable quantities such as luminosity, effective temperature and surface abundances. This has a direct impact on the determinations of derived parameters relying still to a large extent on evolutionary models, as in the case of stellar masses and ages.

The treatment of convectively unstable zones and overshooting regions in evolutionary calculations severely affects the interior structure of the stellar models. This depends on how the boundaries of the convective regions are defined and convective zones themselves treated from the point of view of mixing of elements. The resulting luminosity, temperature, and main sequence lifetime of the star are largely affected by these processes. In the case of massive star evolution, one of the mixing processes that has been extensively studied is semiconvection. During the hydrogen-burning phase the convective core retreats leaving behind a non-uniform chemical profile determined by the composition of the core at the moment each layer is detached from it. Depending on the criterion considered for defining convective boundaries, these layers could maintain their chemical composition or undergo some mixing process, known as semiconvective mixing. Since the pioneering work of \citet{sh58}, several authors have investigated the occurrence of semiconvective mixing and its effects on stellar evolution \citep[e.g.][]{rs70,sc75,cc78,nl85}. Although it was initially thought to occur only in massive stars, semiconvection was also predicted to take place in low-mass stars \citep[e.g.][]{rm72,fc73,hs75,gn77} but with a larger impact during the helium-burning phase.

During the main sequence evolution, stars with masses larger than $\sim$1.1 M$_{\sun}$ develop a convective core whose extent is determined by their temperature stratification. As the timescale for mixing of elements in this convective region is much shorter than the nuclear timescale, the core is believed to be homogeneously mixed and a discontinuity in density appears at the edge of the fully mixed core. This discontinuity is produced either by convective core expansion due to the increasing importance of the \textit{CNO} cycle over the \textit{pp} chain \citep{rm72,hs75}, or by the retreating convective core leaving behind a non-uniform chemical profile \citep{fc73}. Both cases produce higher opacities outside the convective core and allow a semiconvective region to develop. If the resulting density barrier is sustained throughout the main sequence evolution the homogeneously mixed central region will be either restricted in its growth or not allowed to develop at all, drastically changing the behavior of the evolutionary tracks especially close to the main sequence turn-off. This has important implications for the use of the Color-Magnitude Diagram (CMD) morphology and the existence of a hook-like feature at the turn-off to estimate properties of stars at the end of the main sequence and the age of a given stellar population \citep[see for instance][]{am74a,am74b,mm91}.

While it is not possible to obtain direct information about stellar interiors, helioseismology has provided the best example of indirect observations from the interior of a star by piercing the outer layers of the Sun using the oscillations observed at its surface. These data have served to extract the adiabatic sound speed and density profiles of the Sun, determine the location of the solar helium second ionization zone and the base of the convective envelope, as well as to constrain the solar surface abundances \citep[see][and references therein]{jcd02,ba08}. With this principle in mind, efforts are being addressed to extend the techniques applied to the Sun to other stars by means of asteroseismology, in the light of the data currently being obtained by the CoRoT \citep{ab06} and \textit{Kepler} \citep{wb09a} missions.

We address in this paper the issue of the determination of convective boundaries and the treatment of zones which present a gradient in the molecular weight by the inclusion of the Ledoux criterion for convective instability, and a diffusive approach for semiconvective mixing in our stellar evolution code. We explore the possibility of determining the presence of a convective core and discriminating different sizes of homogeneously mixed central regions and evolutionary stages by means of asteroseismology. We do this by exploring the effects that different types of convective mixing processes have on asteroseismic frequency combinations, in agreement with the expected quality of data being obtained by the aforementioned space missions.

The paper is organized as follows. In Sect.~\ref{s:convec} we describe the differences in the definition of the convective boundaries for each stability criterion, and the prescription for mixing in convective, semiconvective and overshooting regions. Sect.~\ref{s:semiconv} focuses in the impact of these boundary definitions and mixing processes during the main sequence evolution. In Sect.~\ref{s:ast} we discuss the method of isolating the influence of the stellar core by means of different asteroseismic tools. We assess the possibility of disentangling between the different mixing prescriptions using the expected observable range of frequencies. A discussion on the open issues and detection possibilities is given in Sect.~\ref{s:disc}. We conclude and state final remarks in Sect.~\ref{s:conc}.
\section{Stellar Modeling}\label{s:convec}
We used the Garching Stellar Evolution Code \citep[GARSTEC,][]{ws08} for our calculations. The interested reader can find in that reference a more detailed description of the numerics and input physics included in the code. For our model calculations we used the 2005 version of the OPAL equation of state \citep{fr96,rn02} complemented with the MHD equation of state for low temperatures \citep{hm88}, low temperature opacities from \citet{jf05} and OPAL opacities for high temperatures \citep{ir96}, the \citet{gs98} solar mixture, and the NACRE compilation for thermonuclear reaction rates \citep{ca99}. Convective zones are treated with the mixing-length theory (MLT), no matter which criterion is used to define their boundaries. Within these zones, the chemical composition is modified either instantaneously or by a diffusive process using the convective velocity estimated from the MLT as described, for instance, in \citet{kw90}.

For the frequency computations of the specific models analyzed, we have used the Aarhus adiabatic oscillation package \citep[ADIPLS,][]{jcd08}.
\subsection{Boundaries of convective and semiconvective zones}\label{ss:cases}
The convective zones within a stellar model are defined using a stability criterion. The convectively unstable regions are usually determined by the Schwarzschild criterion \citep{sh58} by comparing the radiative and adiabatic temperature gradients,  denoted $\nabla_{\mathrm{rad}}$ and $\nabla_{\mathrm{ad}}$ respectively. A layer is convective when
\begin{equation}
\nabla_{\mathrm{ad}} < \nabla_{\mathrm{rad}}.
\end{equation}
However, gradients of molecular weight $\mu$ can have a stabilizing -- or destabilizing -- effect on the convection process. The Ledoux criterion \citep{pl47} takes into account this effect. Within this prescription, a layer is convective when 
\begin{equation}
\nabla_{\mathrm{L}} < \nabla_{\mathrm{rad}},
\end{equation}
with $\nabla_{\mathrm{L}}$, the Ledoux temperature gradient, defined as
\begin{equation}\label{eqn:led_ori}
\nabla_{\mathrm{L}}=\nabla_{\mathrm{ad}}+\frac{\varphi}{\delta}\nabla_{\mu},
\end{equation}
where
\begin{equation}\label{eqn:partials}
\varphi = \left(\frac{\partial\ln\rho}{\partial\ln\mu}\right)_{\mathrm{P}, T}, \delta = -\left(\frac{\partial\ln\rho}{\partial\ln T}\right)_{\mathrm{P, \mu}}, \nabla_{\mu}= \left(\frac{d\ln\mu}{d\ln P}\right).
\end{equation}
For an equation of state appropriate to a mixture of an ideal gas and black body radiation, Eq.~\ref{eqn:led_ori} reduces to:
\begin{equation}\label{eqn:led}
\nabla_{\mathrm{L}}=\nabla_{\mathrm{ad}}+\frac{\beta}{4-3\beta}\nabla_{\mu},
\end{equation}
where $\beta$ is the ratio of gas pressure to total pressure \citep{kw90}.

The inclusion of changes in the molecular weight in the definition of the relevant temperature gradients leads to the appearance of zones whose energy transport process will depend on the convective criterion considered. There are regions which would be considered convective if the Schwarzschild criterion were used, but are in turn radiative if the Ledoux criterion were applied. These zones are called semiconvective zones, and are defined as the regions where
\begin{equation}\label{eqn:sc}
\nabla_{\mathrm{ad}} < \nabla_{\mathrm{rad}} < \nabla_{\mathrm{L}}\,.
\end{equation}
\begin{figure}[!ht]
\centering
\includegraphics[width=\linewidth]{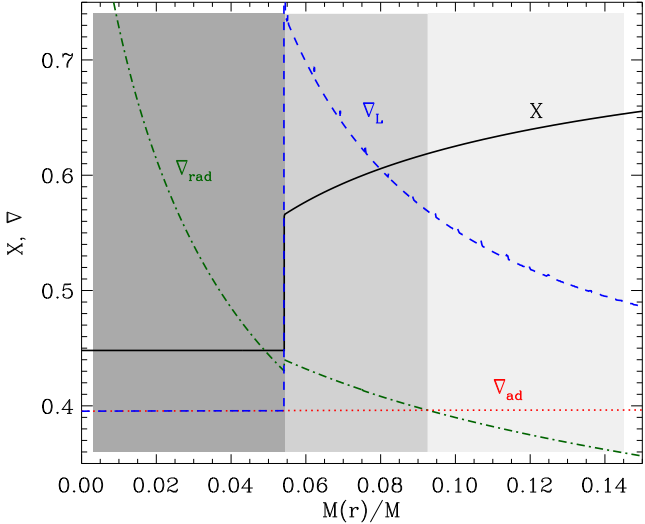}
\caption{Internal structure of a  1.5 M$_{\sun}$ stellar model in the main sequence. The hydrogen abundance profile X (black solid line), the adiabatic gradient $\nabla_{\mathrm{ad}}$ (red dotted line), the radiative gradient $\nabla_{\mathrm{rad}}$ (green dash-dotted line), and the Ledoux gradient $\nabla_{\mathrm{L}}$ (blue dashed line) are plotted as a function of the mass fraction. From left to right, tones of gray fill regions of different energy transport processes: convective zone, semiconvective zone, and radiative zone.}
\label{Grad}
\end{figure}
As an example, a model of a 1.5 M$_{\sun}$ star displaying a semiconvective zone during its main sequence evolution is depicted in Fig.~\ref{Grad}. The hydrogen profile in the stellar interior is constant inside the homogeneously mixed convective core, and presents a sharp variation at the position where the edge of the core is located. This feature produces both a step in the radiative gradient $\nabla_{\mathrm{rad}}$ due to the change in the opacities, and the departure of the Ledoux gradient $\nabla_{\mathrm{L}}$ from the adiabatic gradient $\nabla_{\mathrm{ad}}$ as a result of the change in the molecular weight. Therefore, a significant semiconvective layer appears at the boundary of the convective core. We notice that, in this example, the semiconvective zone is 60\% as massive as the convective region itself. If the Schwarzschild criterion were used instead, the semiconvective layer would be part of the convective core.
\subsection{Prescription for semiconvective layers}\label{ss:mixing}
Throughout the years, several approaches have been proposed to deal with semiconvective regions when they appear in stellar models \citep{ds79,nl83,hs92,gt96}. According to the used prescription, the semiconvective zone is considered to be mixed more or less efficiently \citep[see the comparison made by][]{wm95}. The amount of mixing in semiconvective layers is crucial: an efficient mixing process can reduce the molecular weight gradient sufficiently, and then trigger convective instability. For the present work, we have implemented in GARSTEC the prescription proposed by \citet{nl83,nl85}, based on the description of vibrational instability made by \citet{sk66}. With this method, the true temperature gradient $\nabla$ in a semiconvective region is calculated from the relation
\begin{equation}\label{eqn:real}
\frac{L_{\mathrm{sc}}}{L_{\mathrm{rad}}}=\alpha_{\mathrm{sc}} \frac{\nabla-\nabla_{\mathrm{ad}}}{2\nabla\left(\nabla_{\mathrm{L}}-\nabla\right)}\left[(\nabla-\nabla_{\mathrm{ad}})-\frac{\beta\left(8-3\beta\right)}{32-24\beta-\beta^{2}}\nabla_{\mathrm{\mu}}\right]\,,
\end{equation}
where $L_{\mathrm{sc}}$ and  $L_{\mathrm{rad}}$ are the semiconvective and radiative luminosities, respectively.

The mixing is treated as a time-dependent diffusive process, with the diffusion coefficient calculated as
\begin{equation}\label{eqn:dif_cons}
D_{\mathrm{sc}}=\alpha_{\mathrm{sc}} \frac{\kappa_\mathrm{r}}{6\ c_{\mathrm{p}}\ \rho} \frac{\nabla-\nabla_{\mathrm{ad}}}{\nabla_{\mathrm{L}}-\nabla}\,,
\end{equation}
where $\kappa_{\mathrm{r}}$ is the radiative conductivity, $c_{\mathrm{p}}$ the specific heat at constant pressure, and $\rho$ the density. Both the diffusion coefficient and the temperature gradient in this prescription depend on an efficiency parameter of semiconvection ($\alpha_{\mathrm{sc}}$). The meaning of this parameter is further discussed in Sect.~\ref{ss:mainseq}. We stress the point that this diffusive and time-dependent approach significantly differs from the ones used in previous efforts to treat semiconvection in low-mass stars. In previous studies, the mixing in the semiconvective layer is either performed by adjusting the composition until convective neutrality according to the Schwarzschild criterion is again reached \citep{cm82,am08}, or approximating the results obtained by \citet{hs92} with a two-step function \citep{pd05}.
\subsection{Overshooting}\label{ss:ov}
In some of our models we have considered mixing of chemical elements beyond the formal convective boundaries. We assume that this process induces mixing but does not modify the thermal structure of the layers \citep[\textit{overshooting}, if we use the terminology given by][]{jpz91}. It is implemented in our code as a diffusive process consisting of an exponential decline of the convective velocities within the radiative zone \citep{bf96}. The diffusion constant is given by
\begin{equation}\label{eqn:ove}
 D_{\mathrm{ov}}\left(z\right) = D_0 \ \exp \ \left(\frac{-2z}{\xi H_p}\right)\,,
\end{equation}
where $\xi$ corresponds to an efficiency parameter, $H_p$ is the pressure scale height, $z$ is the distance from the convective border, and the diffusion constant $D_0$ is derived from MLT-convective velocities \citep{kw90}. We have used an overshooting efficiency of $\xi=0.016$, which is in the range expected for main sequence stars \citep{fh97} and corresponds to the value obtained by calibrating the parameter $\xi$ with open clusters. The size of the overshooting region is further limited in the case of small convective cores. This is done in GARSTEC using a geometrical cutoff factor, allowing the overshooting region to extend only to a fraction of the convective zone \citep[see][for details]{ws08,zm10}.
\section{Impact on the Stellar Models}\label{s:semiconv}
Several models were computed using the different mixing prescriptions and definition of the convective boundaries. Within a convective zone the mixing is performed diffusively using the MLT while in a semiconvective region the mixing is carried out as explained in Sect.~\ref{ss:mixing}. Calculations were made for both the Schwarzschild and the Ledoux criterion for the definition of the convective zones, with and without including extra mixing due to overshooting and semiconvection. All the models considered here are computed for solar metallicity. We explore the effects of these processes for masses ranging from 1 to 2 M$_{\sun}$ starting at the pre-Zero Age main sequence and evolved until hydrogen exhaustion in the core.

\subsection{Convective boundary definition criteria}\label{ss:criteria}
Applying either the Schwarzschild or the Ledoux criterion influences the appearance and the size of convective cores as a function of stellar mass. To study these effects, we computed models with no convective overshoot and no mixing throughout the semiconvective regions when these were present ($\alpha_{\mathrm{sc}}=0$).

\begin{figure}[!ht]
\centering
\includegraphics[width=\linewidth]{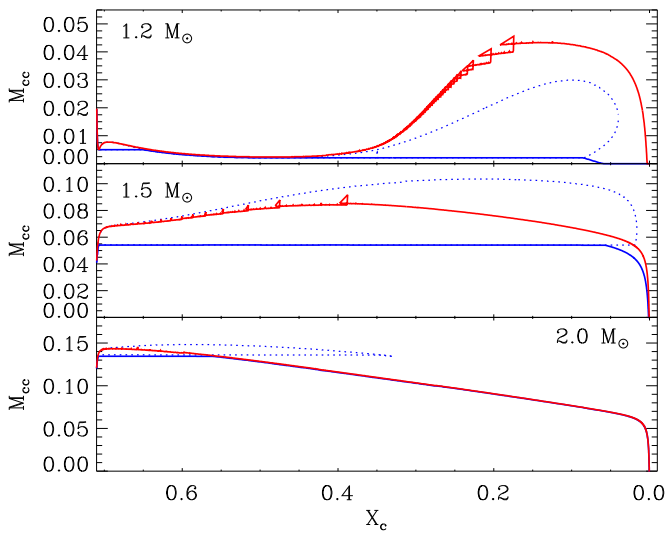}
\caption{Evolution of the convective core size in mass coordinate as a function of central hydrogen content during the main sequence phase for the two considered convective boundary definition criteria: Ledoux (solid blue lines) and Schwarzschild (solid red lines). Blue dotted lines depict semiconvective zones when the Ledoux criterion is applied (by definition there are no semiconvective zones when the Schwarzschild criterion is applied). The panels represent different masses: 1.2 M$_{\sun}$ (\textit{top}), 1.5 M$_{\sun}$ (\textit{middle}) and 2.0 M$_{\sun}$ (\textit{bottom}).}
\label{Ccore1}
\end{figure}

In Fig.~\ref{Ccore1} we present the evolution during the main sequence of convective and semiconvective regions in the interiors of three representative models:
\begin{itemize}
\item a model where the inclusion of the Ledoux criterion inhibits the growth of the convective core in the late phase of the main sequence evolution (1.2 M$_{\sun}$ model, Fig.~\ref{Ccore1}~\textit{top panel});
\item  a model with a convective core increasing in size during the hydrogen burning phase whose growth is constrained if the Ledoux criterion is applied (1.5 M$_{\sun}$ model, Fig.~\ref{Ccore1}~\textit{middle panel});
\item a model where the convective core recedes during the main sequence leaving behind a chemical discontinuity (2.0 M$_{\sun}$ model, Fig.~\ref{Ccore1}~\textit{bottom panel}).
\end{itemize}
There are clear differences in the convective core behavior for the two considered criteria. In the case of growing convective cores (Fig.~\ref{Ccore1} top and middle panels) the molecular weight discontinuity at the edge of the convective zone produces a steep increase in the opacity, which translates into a sharp discontinuity in the radiative gradient (as shown in Fig.~\ref{Grad}).

If the Schwarzschild criterion is applied, the profile of the convective core presents wrinkles throughout the evolution, signature of a convective process not properly taken into account \citep{yl08}. The exact position of the convective core edge will depend largely in the numerics of the evolutionary code and the way the mesh points are placed where the discontinuity in the molecular weight is located \citep{am08}. Slight variations in the radiative gradient are induced by, for instance, the interpolation in the opacity tables, the allowed time-step and the accuracy of the calculations, resulting in a small region right outside the core edge becoming convective and supplying fresh hydrogen to the outer layer of the core. The core expands and produces the crumple profiles (backwards loops) observed in the top and middle panels of Fig.~\ref{Ccore1}, in a phenomena similar to the so-called \textit{breathing pulses} in horizontal branch stars \citep[see for example][for a detailed discussion and further references on the topic]{mc07}.

If the Ledoux criterion is applied instead, the convective core profile remains flat and continuous throughout the main sequence evolution, as the molecular weight discontinuity at its edge inhibits its growth. The semiconvective zone appearing outside the core is not mixed through and thus becomes larger during the hydrogen burning phase. At the end of the main sequence evolution, the convective core recedes leaving a radiative layer between it and the semiconvective zone, causing a change in the temperature gradients and making the semiconvective region to ultimately disappear.

In a model with a receding core the convective central region reaches its largest size during the initial phase of hydrogen-burning, the Schwarzschild criterion allowing it to grow to a slightly larger extent than the Ledoux criterion (Fig.~\ref{Ccore1}, bottom panel). The chemical discontinuity left behind by the retreating core leaves an imprint in the molecular weight which permits a semiconvective zone to develop, but the radiative gradient decreases following the shrinking of the core transforming the zone into a radiative one. Since the effect is almost negligible for this case, we will focus for the rest of the paper on the models where the inclusion of the Ledoux criterion plays an important role (masses below $\sim$1.7 M$_{\sun}$ with growing convective cores).
\subsection{Main sequence evolution}\label{ss:mainseq}
In the previous section we have shown the impact that different mixing prescriptions have on the size of the homogeneously mixed central region during the main sequence evolution. The size (and existence) of the convective core will determine the appearance (or lack) of a hook-like feature in the CMD at the end of the hydrogen-burning phase. This feature at the turn-off is used in the determination of the age of stellar populations by fitting isochrones reproducing the shape of the hook at a certain metallicity, thus setting constraints on the stellar mass at the end of the main sequence for a given set of input physics \citep[as an example, see][for the specific case of M67]{dv07,zm10}. As discussed in Sect.~\ref{ss:criteria}, using the Ledoux criterion can inhibit the growth of the convective core for the 1.2 M$_{\sun}$ model and therefore the appearance of the hook feature in the evolutionary track. This is the case when no mixing is performed within the developing semiconvective region ($\alpha_{\mathrm{sc}}=0$) and the molecular weight barrier caused by hydrogen burning does not allow the core to grow.

\begin{figure}[!ht]
\centering
\includegraphics[width=\linewidth]{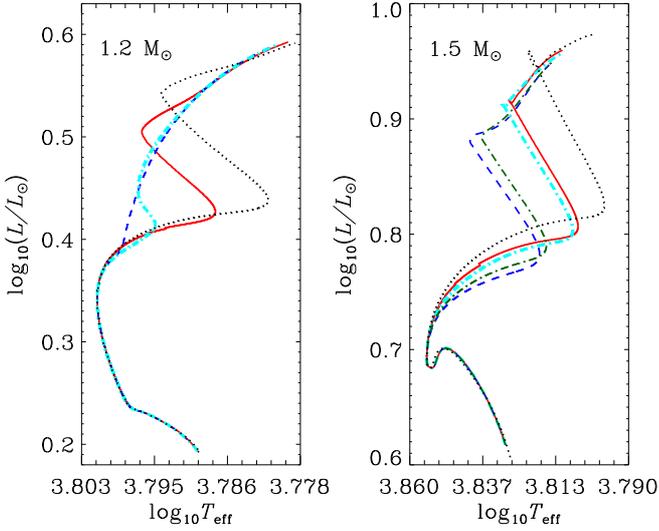}
\caption{Hertzprung-Russell Diagram for 1.2 M$_{\sun}$ (\textit{left panel}) and 1.5 M$_{\sun}$ (\textit{right panel}) models. Computations done with the Ledoux criterion and no mixing in the semiconvective regions are plotted with dashed blue lines, Schwarzschild criterion with solid red lines, and dotted black lines correspond to tracks with Schwarzschild criterion and overshooting. Two values for the semiconvective efficiency $\alpha_\mathrm{sc}$ are also plotted in thin green dash-dotted line ($\alpha_{\mathrm{sc}} = 0.001$; not shown for the 1.2 M$_{\sun}$ as it overlaps with the dashed track) and thick light blue dash-dotted line ($\alpha_{\mathrm{sc}} = 0.01$). The tracks span the evolution from the Zero-Age main sequence until hydrogen exhaustion in the core.}
\label{hrdms}
\end{figure}

The amount of semiconvective mixing is controlled by the efficiency parameter $\alpha_{\mathrm{sc}}$ and has a value restricted to $\alpha_{\mathrm{sc}} < 1$, but has been used usually in the range $0.001 < \alpha_{\mathrm{sc}} < 0.1$ \citep{nl85,nl91,wm95}. Higher values of $\alpha_{\mathrm{sc}}$ mean faster mixing velocities, achieving instantaneous mixing when $\alpha_{\mathrm{sc}}\rightarrow\infty$. In Fig.~\ref{hrdms} we have plotted evolutionary tracks for 1.2 and 1.5 M$_{\sun}$ cases calculated with different mixing prescriptions and semiconvective efficiency parameters. By increasing the value of $\alpha_{\mathrm{sc}}$ the convective core develops to a larger extent and the evolutionary tracks resemble the one calculated with the Schwarzschild criterion. For the 1.2 M$_{\sun}$ case the track computed with $\alpha_{\mathrm{sc}} = 0.001$ does not increase the size of the convective core with respect to the track calculated with no mixing in the semiconvective zone, thus their tracks overlap in the Hertzprung-Russel Diagram (HRD). One interesting aspect is that a position in the HRD can be shared by stars with the same mass but different evolutionary stages, internal structures and stellar ages.

\begin{figure}[!ht]
\centering
\includegraphics[width=\linewidth]{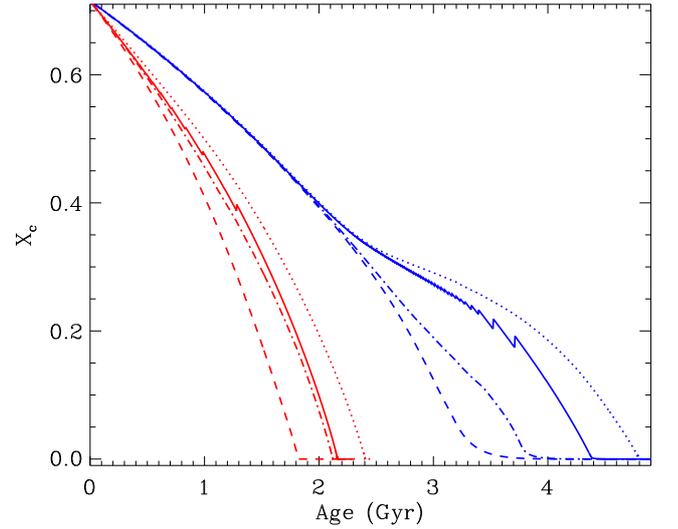}
\caption{Central hydrogen content of the selected models during their main sequence evolution: 1.2 M$_{\sun}$ (blue) and 1.5 M$_{\sun}$ (red). \textit{Solid lines}: Schwarzschild criterion without overshooting.  \textit{Dashed lines}: Ledoux criterion with no semiconvective mixing ($\alpha_{\mathrm{sc}} = 0$). \textit{Dash-dotted lines}: Ledoux criterion and a semiconvective efficiency of $\alpha_{\mathrm{sc}} = 0.01$. \textit{Dotted lines}: models with convective overshooting.}
\label{age}
\end{figure}

Changing the prescription for convection has an impact on the stellar age at which the end of the main sequence is reached. In the diffusive prescription we use for semiconvective zones, $\alpha_{\mathrm{sc}}$ modifies the value of the diffusion coefficient and also the true temperature gradient in the semiconvective zone. The growth of the convective core is not enhanced but restricted due to the molecular weight gradient and no extra fuel supply is added to the core, which translates into shorter main sequence lifetimes. In Fig.~\ref{age} we present the hydrogen contents at the center of the models as a function of age, where the differences are clearly visible for the considered mixing prescriptions and can amount to 30\% of the main sequence lifetimes.
\subsection{Effects of mixing in semiconvective layers}\label{ss:alpha} 
\begin{figure}[!ht]
\centering
\includegraphics[width=\linewidth]{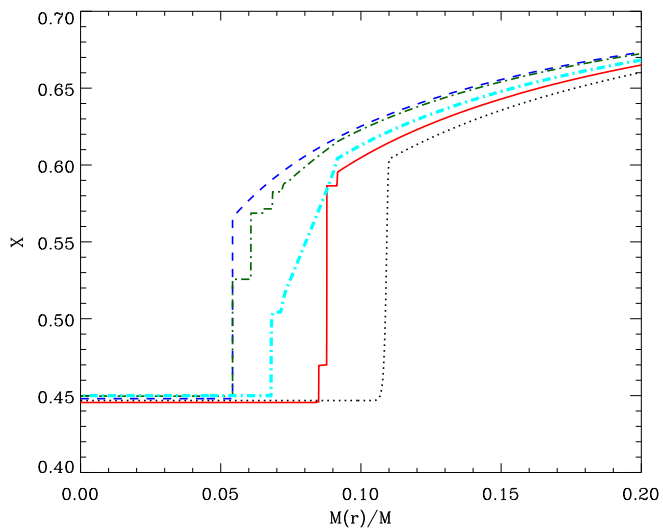}
\caption{Hydrogen profiles for similar central hydrogen contents of the 1.5 M$_{\sun}$ models and different mixing prescriptions: Ledoux criterion and no mixing in the semiconvective zones (dashed blue line), $\alpha_{\mathrm{sc}} = 0.001$ (thin dash-dotted green line), $\alpha_{\mathrm{sc}} = 0.01$ (thick dash-dotted light blue line), Schwarzschild criterion (solid red line) and one model including convective overshooting (dotted black line).}
\label{chem_prof}
\end{figure}

In Fig.~\ref{chem_prof} we present hydrogen profiles near the center for 1.5 M$_{\sun}$ models calculated with the Ledoux criterion and different semiconvective mixing efficiency values, with the Schwarzschild criterion and with overshooting. The sizes of the homogeneously mixed zones are different as well as the shape of the chemical profile at the edge of the convective core. 
It is worth noticing that for the 1.5 M$_{\sun}$ case considered in this paper, a semiconvective coefficient of $\alpha_{\mathrm{sc}}\sim 0.01$ is already efficient enough to very closely reproduce the results obtained when the Schwarzschild criterion is applied (see the evolutionary tracks in Fig.~\ref{hrdms}). The same is true for $\alpha_{\mathrm{sc}}\sim 0.1$ in the 1.2 M$_{\sun}$ model. This suggests that the value of the semiconvective efficiency required to reproduce the results obtained with the Schwarzschild criterion decreases with mass. In relative terms, the size of the semiconvective zone compared to the total convective core size is smaller at higher masses, leading to a shorter mixing timescale.

Convective overshooting has for long been thought of as a natural process capable of smoothing out molecular weight gradients close to the convective core \citep[e.g.][]{am74a,am74b,nl91,an10}. We computed models using the Ledoux criterion and the overshooting prescription described in Sect.~\ref{ss:ov} and realized that this is also the case for low-mass stars using our calibrated value for the overshooting efficiency. One should keep in mind, though, that the calibration itself was based on certain assumptions about convection \citep[e.g][]{ap04,dv06}.
\section{Asteroseismic diagnoses}\label{s:ast}
We have highlighted in Sect.~\ref{s:semiconv} the importance of the different mixing prescriptions considered in this paper on structure and evolution of low-mass stars. In practical terms, these convective processes translate into different convective core sizes, ages, and evolutionary stages for compatible positions in the HRD. The quest of detecting the variations in stellar interiors produced by differences in the mixing and energy transport processes can be fulfilled by a technique capable of piercing the outer layers of stars and which is sensitive to density discontinuities. In this respect, asteroseismology is thought to be the most powerful tool to study the inner properties of stars through the observation of stellar oscillations \citep[e.g][and references therein]{mcu07,ca08,jcd10a}.

Many efforts have been conducted to discover the physics of stellar interiors by understanding the influence that different physical phenomena have on the observed frequency spectra. In particular, it is known that a discontinuity in the chemical profile of a star produces a sharp variation in the adiabatic sound speed, which in turn introduces an oscillatory component in the frequencies \citep{sv88,dg90,jp93}. This fact has already been applied to determine the position of the base of the convective envelope in the Sun \citep{jcd91}, to constrain the amount of overshoot below it \citep{mm94,rv94c,jcd95}, and to constrain the properties of the solar helium second ionization zone \citep{jcdph91,ab94,mmmt05}. The same technique has been proposed to determine the position of the base of the convective envelope and estimate the helium abundance for stars other than the Sun \citep{mm00,mmmt98}.

Following this principle, convective cores in low-mass stars have been the subject of several studies by means of asteroseismology aiming to determine the evolutionary state and size of the mixed central region in stars \citep[e.g][]{ap94,na95,ma01,am06,cm07}. Most of these investigations have focused their attention on constraining the size and chemical composition of the convective core, while only a few of them have acknowledged the impact of its boundary definition and extra mixing processes such as semiconvection and overshooting in the oscillation frequencies \citep{pd05,mg07,am08,yl09}. In this section we aim at finding a seismic tool that isolates the stellar core and which is sensitive to its size and central hydrogen content, allowing to disentangle between the different mixing prescriptions.
\subsection{Seismic variables suited to isolate the core}\label{ss:ratios_def}
In order to study the inner structure of a star using asteroseismology, first we need to find an appropriate seismic variable which allows us to probe the desired region of the star and extract the required information. Different combinations of low-degree p-modes have been suggested as suitable probes of the physical characteristics of a star \citep[e.g.][]{jcd84}, the most commonly used being the so-called large and small frequency separations defined as:
\begin{eqnarray}\label{eqn:diff}
\Delta_{l}(n) & = & \nu_{n,l}-\nu_{n-1,l}\\
d_{l,l+2}(n) & = & \nu_{n,l}-\nu_{n-1,l+2}\,,
\end{eqnarray}
where $\nu_{n,l}$ is the mode frequency of angular degree $l$ and radial order $n$. These combinations are affected by the outer layers of the star where turbulence in the near-surface is almost never taken into account into star modeling \citep[e.g.][]{jb04}.

In our case, we need to isolate from surface contamination the signal arising from the interior of the star in order to properly quantify the effects in the frequency spectra of the presence of a convective core. \citet{rv03a} proposed to use the smooth 5 points small frequency separations and the ratio of small to large separations, and showed that these quantities are mainly determined by the inner structure of the star. They are constructed as:
\begin{equation}\label{eqn:d01}
d_{01}(n)=\frac{1}{8}(\nu_{n-1,0}-4\nu_{n-1,1}+6\nu_{n,0}-4\nu_{n,1}+\nu_{n+1,0})
\end{equation}
\begin{equation}\label{eqn:d10}
d_{10}(n)=-\frac{1}{8}(\nu_{n-1,1}-4\nu_{n,0}+6\nu_{n,1}-4\nu_{n+1,0}+\nu_{n+1,1})
\end{equation}
\begin{eqnarray}\label{eqn:rat}
r_{01}(n)=\frac{d_{01}(n)}{\Delta_{1}(n)},& & r_{10} = \frac{d_{10}(n)}{\Delta_{0}(n+1)}\,.
\end{eqnarray}
The small frequency separations have already been used to identify the location of the convective envelope and the helium second ionization zone in the Sun by applying them to observational data \citep{ir09}, while it has been shown that the ratios fairly cancel out the influence in the frequencies of the outer layers \citep{ir04,ir05,of05}. As mentioned before, sharp variations on the adiabatic sound speed produce an oscillatory signal in the p-mode spectra whose period is related to the position of such a discontinuity. In particular, the period of this oscillation relates to the travel time of the wave through the acoustic cavity, and therefore to the radial coordinate where the discontinuity is located \citep{mm94,rv94a,rv01}. This travel time is represented by the acoustic radius, given by:
\begin{eqnarray}\label{eqn:time}
t & = & \int_{0}^{r}\frac{\mathrm{d}r}{c_s}\,
\end{eqnarray}
where $c_s$ is the adiabatic sound speed. There is an alternative representation of the acoustic radius called the acoustic depth ($\tau$), which measures the travel time from the surface towards the interior. Therefore, if the total acoustic radius is given by $\tau_c = t(R)$, then clearly $\tau = \tau_c - t$. If the location of the density discontinuity is given by, say, $r_1$ in radial coordinates, and that same position is represented in acoustic radius and acoustic depth by $t_1$ and $\tau_1$ respectively, the periods of the oscillation in frequency space induced by this sharp variation are 1/(2$t_1$) and 1/(2$\tau_1$), according to the considered variable. 
That being the case, if either the ratios or the small separations are affected only by the position of the convective core, we could extract this information from the frequency data.
\subsection{Sensitivity of $r_{10}$ and $r_{01}$ to the core}\label{ss:rat_examp}
\begin{figure*}[!t]
\centering
\includegraphics{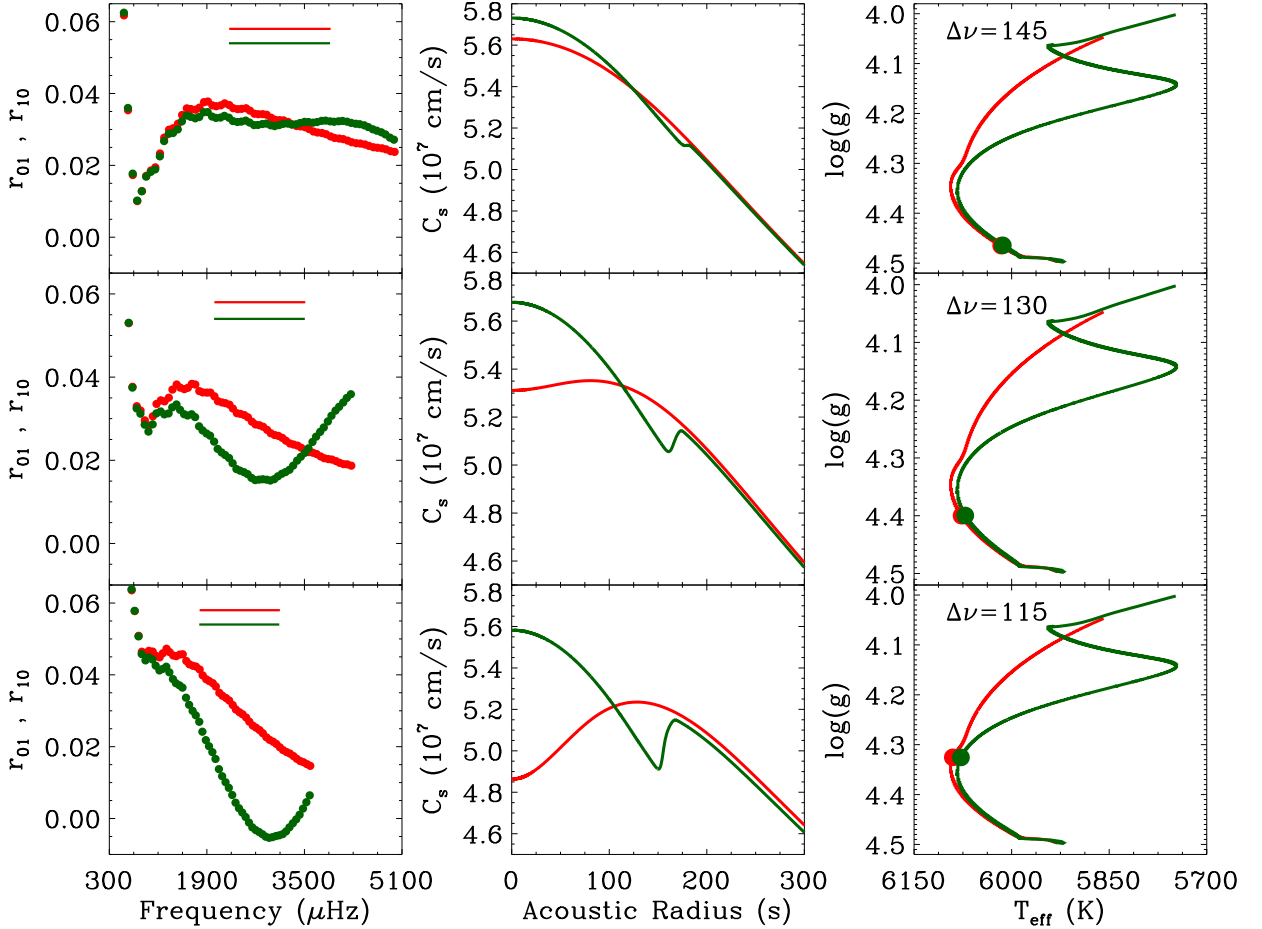}
\caption{1.1 M$_{\sun}$ models with a convective core (green) and without one (red). Each row depicts models sharing the same large frequency separation. \textit{Left Panels} frequency ratios as defined in Eq.~\ref{eqn:rat}. Horizontal lines show a typical observable frequency range for each mixing prescription of radial orders n=15-27. \textit{Central Panels:} adiabatic sound speed as a function of the acoustic radius. \textit{Right Panels:} evolutionary tracks for the two cases considered, the position of the studied models being represented by a filled circle. Values of the selected large frequency separations are given (in $\mu$Hz)}
\label{core_comp}
\end{figure*}
If the differences and ratios defined in Eqs.~\ref{eqn:d01}, \ref{eqn:d10} and \ref{eqn:rat} are indeed sensitive to the deep stellar interior, they should be strongly influenced by the presence of a convective core or a steep gradient in the adiabatic sound speed near the center \citep{rv07}. To test this, we produced two sets of evolutionary tracks for a 1.1 M$_{\sun}$ star, one using the normal Schwarzschild criterion and the other one including overshooting without the geometrical restriction described in Sect. \ref{ss:ov}. This way, we produced a 1.1 M$_{\sun}$ model with a growing convective core during the hydrogen burning phase. In order to compare the effects of the presence of the core, we plotted the evolution during the main sequence of these models in Fig.~\ref{core_comp}. The left panels present the ratios as a function of frequency, while the center panels show the adiabatic sound speed near the center for the selected models, whose position in the observational plane is depicted by a filled circle in the right panels of the figure.

The frequency ratios present a common component in the low frequency domain (see left panels), which therefore should be determined by the global seismic properties of the star: the selected models have the same large frequency separation. At higher frequencies, the ratios deviate from each other revealing the structural differences between the models. This part of the frequency domain presents two contributions: one given by the presence of a convective core and another one dominated by the evolutionary state of the model (equivalent to the central hydrogen content). The oscillatory component observed in the ratio of the model calculated with overshooting is related to the presence of a convective core.

A straight line can be drawn between the point where the ratios start deviating from each other and the frequency value where the oscillatory component of the model with the convective core shows its first minimum. We will refer to this part of the frequency domain as the \textit{linear range} from here on. Independently of the presence of a convective core or not, we observe an increase of the slope -- strictly speaking of its absolute value --  with evolution, suggesting that it traces the central hydrogen content of the star (and hence stellar age). Thus, the sole presence of a negative slope does not ensure the existence of a convective core \citep[see also][]{ib10}.

However, in the last row of panels in Fig.~\ref{core_comp} the models plotted have very different central hydrogen abundances ($X_{c}\sim0.14$ versus $X_{c}\sim0.4$ for the model without and with convective core, respectively). In the model without convective core, the increase of the slope is mainly due to the growth of the density gradient in the center, built during hydrogen burning. Nevertheless, the slope is sensibly higher for the model with a convective core, even if this star is less evolved. This difference is a clue to the presence or absence of a convective core and will be discussed in Sect.~\ref{ss:Pop_plots}.

The global behavior of the small frequency separations is the same as the one observed in the frequency ratios. Thus, we will focus the analysis in the ratios keeping in mind that the same conclusions are applicable to the case of the separations.
\subsection{Influence of mixing prescriptions on $r_{10}$ and $r_{01}$}\label{ss:rat_ms}
\begin{figure*}[!t]
\centering
\includegraphics{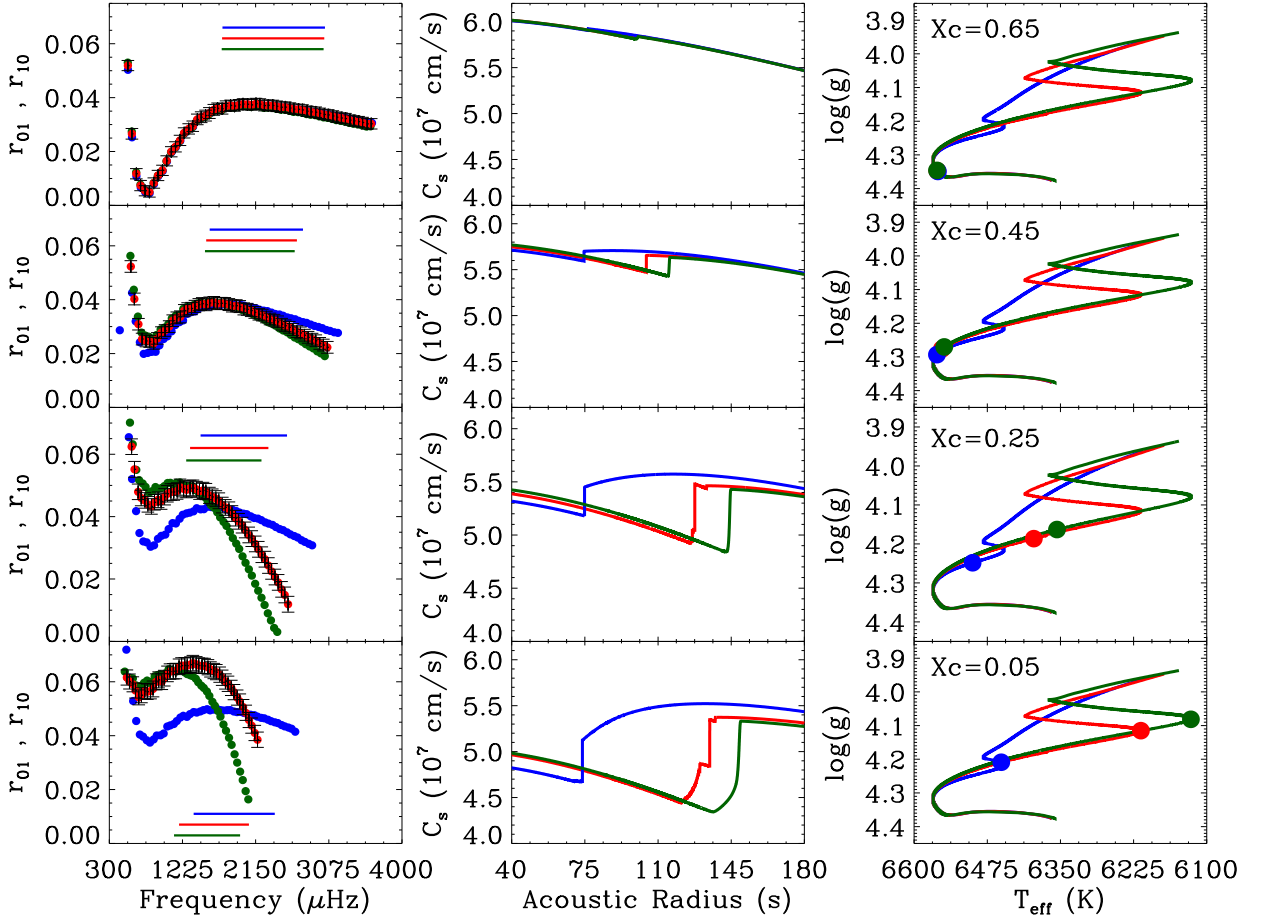}
\caption{Comparison for 1.3 M$_{\sun}$ models using different mixing prescriptions: Ledoux (blue), Schwarzschild (red) and overshooting (green). Each row of models represents a value of the central hydrogen content (from top to bottom): $Xc=0.65$, $Xc=0.45$, $Xc=0.25$, and $Xc=0.05$. \textit{Left Panels:} frequency ratios as defined in Eq.~\ref{eqn:rat}. Horizontal lines show a typical observable frequency range for each mixing prescription of radial orders n=15-27. \textit{Central Panels:} adiabatic sound speed showing the position of the density discontinuity. \textit{Right Panels:} evolutionary tracks for the three selected prescriptions, and the position of the studied models is represented by a filled circle. Error bars have been plotted for reference in the models computed with the Schwarzschild criterion.}
\label{rat_evol}
\end{figure*}
\begin{figure*}[!t]
\centering
\includegraphics{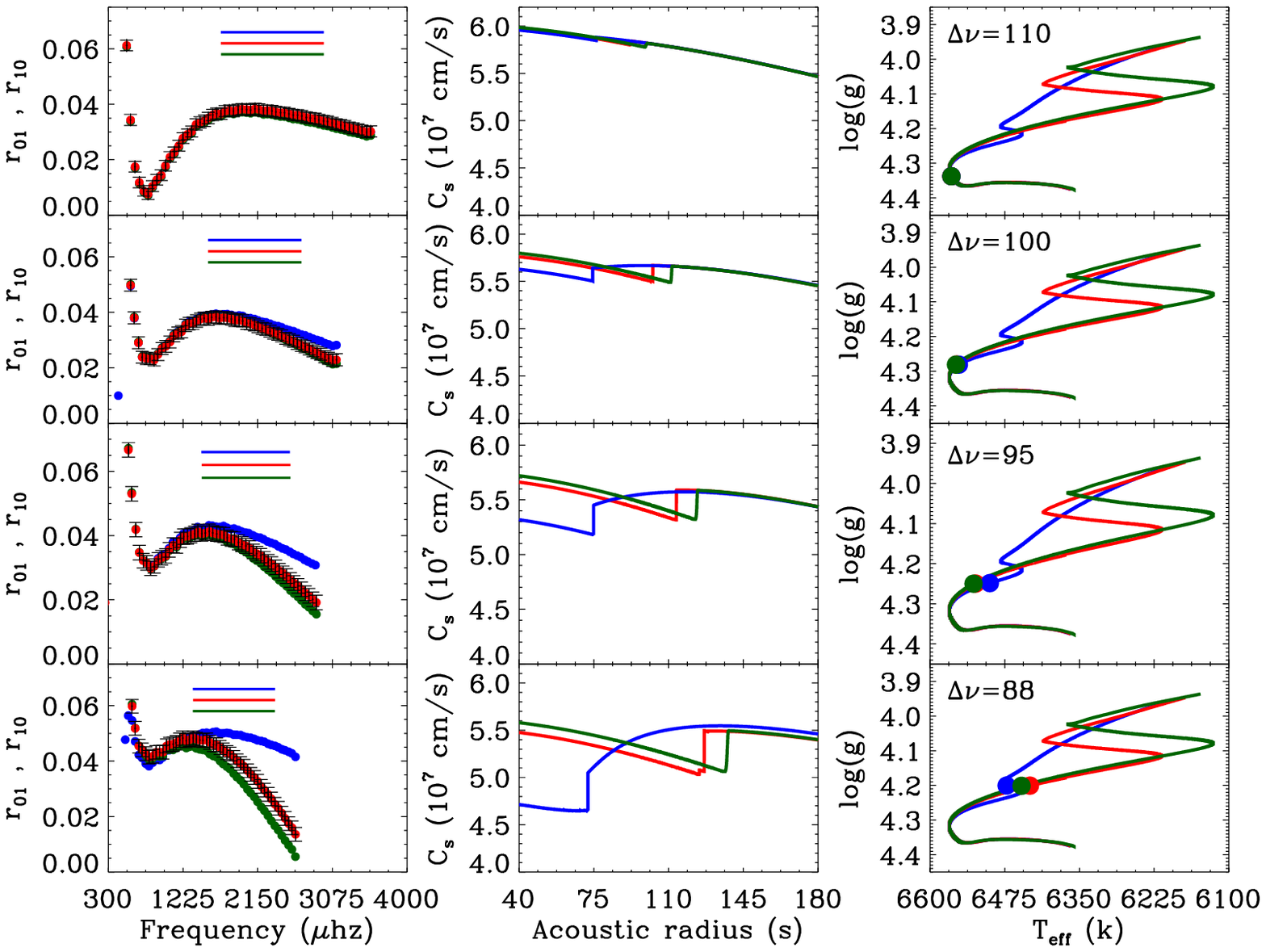}
\caption{Same as Fig.~\ref{rat_evol} but each row represents models with the same large frequency separation value, which is given in the right panel of each row (in $\mu$Hz).}
\label{rat_large}
\end{figure*}
When considering only models which do have a convective core, we can study the sensitivity of the ratios to the position and size of the convective core, and to the size of the density discontinuity. We explore the behavior of 1.3 M$_{\sun}$ models in the three most extreme cases in terms of the size of the mixed region: Ledoux criterion with no extra mixing in the semiconvective region, the normal Schwarzschild criterion, and one evolutionary sequence including overshooting. An error of 0.2 $\mu$Hz in the frequency determinations was assumed, as expected from the CoRoT and \textit{Kepler} missions \citep{ab06,jcd07}.

In Fig.~\ref{rat_evol} we present a similar diagram as in Fig.~\ref{core_comp} showing the changes of this seismic diagnose as a function of the central hydrogen content, mapping therefore the main sequence evolution. When considering one mixing prescription (color) we notice that the absolute value of the slope of the frequency ratios increases with evolution. The considered models have the same central hydrogen content, so when observing this behavior of the ratios nothing can be said about the exact evolutionary stage of the star. However, the size of the discontinuity in the adiabatic sound speed is indeed different, as well as the size of the cores, the largest being the one with the most negative slope.

One is tempted to interpret this as a direct relation between the size of the discontinuity and the absolute slope of the ratios, but two details must be taken into account: for the case of the Ledoux criterion, the change in the slope during the main sequence evolution is much smaller than for the other cases, and the acoustic radius of the fully mixed region remains almost constant (while for the other prescriptions it increases). It is clear that the value of the absolute slope has contributions coming from the position of the convective core and the size of the sound speed discontinuity.

Bearing this in mind, instead of using the central hydrogen content as a proxy for the evolution of the ratios we considered models with the same large frequency separation, as it is the parameter first and most easily obtained from seismic observations. Figure~\ref{rat_large} shows this case for the 1.3 M$_{\sun}$ models with large separations equal to $\Delta\nu=$110, 100, 95 and 88 $\mu$Hz. Due to the differences in the mixing prescriptions, the selected models do not have the same central hydrogen content but almost the same age.

The behavior at the low frequency range ($\nu\lesssim$ 1200 $\mu$Hz) is the same for the three mixing prescriptions for a given large frequency separation value. As already mentioned in Sect.~\ref{ss:rat_examp}, this is an indication of this part of the frequency spectrum being dominated by the global seismic properties of the star, while the frequency values above $\sim$ 1200 $\mu$Hz seem to be sensitive to the differences in the interior (the \textit{linear range}).

Within this frequency region where the information of the core is contained, we can compare in Fig.~\ref{rat_large} a case opposite to that in Fig.~\ref{rat_evol}: the last row presents models with the same size of discontinuity of the adiabatic sound speed ($\sim 0.27\times 10^7$ cm/s). The hydrogen content of the Ledoux (blue) model is only $\sim$ 2$\%$ while for the the Schwarzschild criterion and the model with overshooting it is $\sim$ 28$\%$ and $\sim$ 32$\%$, respectively. The model calculated with the Ledoux criterion again has the smallest absolute slope value and size of the convective core.

We stress the fact that the absolute value of the slope cannot be attributed only to the size of the adiabatic sound speed discontinuity. In fact, we can distinguish three effects directly linked to the global behavior of the ratios:
\begin{itemize}
\item a decrease in the central hydrogen content produces an increase in the slope in the \textit{linear range}, which is visible no matter if a convective core is present or not;
\item when a convective core is present, an increase of the size of the discontinuity in the adiabatic sound speed at the core boundary increases the amplitude of the oscillatory signature, making the slope of the ratios steeper;
\item the extent of the convective core also affects the slope: a larger convective core reduces the period of the oscillatory component in the ratios, increasing in turn the absolute value of the slope.
\end{itemize}
It is not straightforward to determine how much influence on the slope each of these effects have. A differential comparison must be done in order to assess where the major contributions are coming from. Nevertheless, now that the key processes affecting the frequency ratios have been identified, we can use this information to disentangle between models with different central mixed zones sizes and evolutionary stages.
\subsection{Breaking the degeneracy: splitting mixing prescriptions and finding convective cores}\label{ss:Pop_plots}
\begin{table*}
\caption{Main properties of the models for a particular value of the large frequency separation. For each mass value, the structure of the computation for a given mixing prescription is given: Ledoux (Led), Schwarzschild (Sch) and overshooting (Ove). The columns are: effective temperature, gravity (logarithm), stellar age, central hydrogen content, size of the sound speed discontinuity at the edge of the convective core (Jump), position of the edge of the convective core in terms of acoustic radius (Arad$_\mathrm{cc}$), and the slope and mean values of the frequency ratios in the radial order n=15-27.}
\label{table:pop_mass}
\centering
\begin{tabular}{c c c c c c c c c c c}
\hline\hline\\ 
$\Delta\nu (\mu$Hz) & \multicolumn{2}{c}{Model} & T$_\mathrm{{eff}}$ (K) & log(g) & Age (Myr) & X$_\mathrm{c}$ & Jump (10$^7$ $\mathrm{cm\,s^{-1}}$) & Arad$_\mathrm{{cc}}$ (s) & Slope ($\mathrm{\mu Hz^{-1}}$)& Mean\\
\hline\\
96      & 1.10 M$_{\sun}$ & Sch & 6048 & 4.22 & 6130.4 & 1e-5  & ----- & --- & $-23\times 10^{-4}$ & 0.049\\
        &                 & Ove & 5955 & 4.22 & 6531.1 & 0.216 & 0.610 & 150 & $-71\times 10^{-4}$ & 0.007\\
        & 1.25 M$_{\sun}$ & Led & 6379 & 4.22 & 2636.2 & 0.016 & 0.225 & 62 & $-8\times 10^{-4}$ & 0.046\\
        &                 & Sch & 6332 & 4.22 & 2636.2 & 0.245 & 0.434 & 115 & $-21\times 10^{-4}$ & 0.041\\
        &                 & Ove & 6343 & 4.22 & 2630.4 & 0.287 & 0.430 & 125 & $-26\times 10^{-4}$ & 0.034\\
        & 1.30 M$_{\sun}$ & Led & 6463 & 4.23 & 1812.8 & 0.159 & 0.370 & 75 & $-6\times 10^{-4}$ & 0.043\\
        &                 & Sch & 6496 & 4.23 & 1812.8 & 0.347 & 0.338 & 117 & $-15\times 10^{-4}$ & 0.038\\
        &                 & Ove & 6499 & 4.23 & 1810.5 & 0.376 & 0.329 & 128 & $-19\times 10^{-4}$ & 0.033\\
        & 1.35 M$_{\sun}$ & Led & 6628 & 4.24 & 1146.8 & 0.346 & 0.255 & 90 & $-4\times 10^{-4}$ & 0.041\\
        &                 & Sch & 6651 & 4.24 & 1146.8 & 0.444 & 0.213 & 122 & $-10\times 10^{-4}$ & 0.035\\
        &                 & Ove & 6655 & 4.24 & 1144.6 & 0.465 & 0.228 & 131 & $-12\times 10^{-4}$ & 0.034\\
\hline                 
\hline                 
\end{tabular}
\end{table*}
In the examples above we have shown that the frequency ratios are sensitive to the central hydrogen content, the existence and extent of a convective core, and the amplitude of the discontinuity in the adiabatic sound speed. Thus the importance of being able to disentangle between the different mixing prescriptions. An attempt to do so was presented by \citet{pd05}, where they characterized the size of the mixed zone in stellar models using the small frequency separation. The authors studied differences between two types of semiconvective mixing and two values of convective overshooting by averaging over a range of radial orders the small frequency separation and using its slope. We will apply a similar approach to the frequency ratios using the radial orders n=15--27, a typical range which is expected to be detected by the CoRoT and \textit{Kepler} missions (see horizontal lines in the left panels of Figs.~\ref{core_comp}, \ref{rat_evol} and \ref{rat_large}).

\begin{figure}[!ht]
\centering
\includegraphics[width=\linewidth]{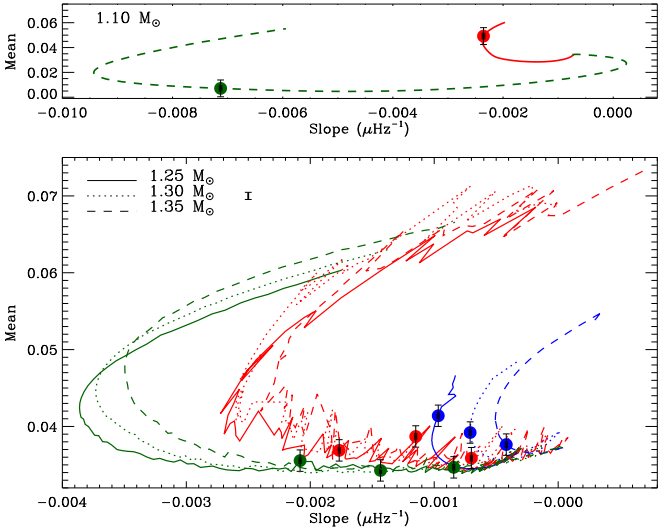}
\caption{Slope versus mean value diagram of the frequency ratios in the assumed observable range throughout the main sequence evolution. \textit{Top panel}: 1.1 M$_{\sun}$ models without overshooting (solid red line) and with overshooting (dashed green line). \textit{Bottom panel}: different mass values for the three considered mixing prescriptions: Ledoux (blue), Schwarzschild (red) and Overshooting (green). A representative 1-$\mathrm{\sigma}$ error bar is plotted on the top left. Models marked with a filled circle have the same large frequency separation value, and error bars correspond to 3-$\mathrm{\sigma}$ uncertainties in the frequencies. See text for details.}
\label{popMass}
\end{figure}

In Fig.~\ref{popMass} we present a slope versus mean diagram in the observable range of the frequency ratios, for the complete main sequence evolution of different masses and mixing prescriptions. For each mass value and mixing prescription we have also selected one model with a large frequency separation value of 96 $\mu$Hz, their properties presented in Table~\ref{table:pop_mass}. The position of the zero-age main sequence model for a given mass value is the same independent of the applied mixing prescription, as the interiors are until that point the same. Once the models start evolving, the different behaviors in their frequency ratios make them clearly distinguishable. The tracks reach a minimum value in slope near the exhaustion of the hydrogen in the center (when the convective core starts receding if present).

For the 1.1 M$_{\sun}$ tracks the two cases present very different values of slopes throughout the evolution, allowing us to disentangle between a star with a convective core and one that has a radiative interior. This is particularly interesting as the stellar parameters are almost the same for both stars, so their positions in the CMD overlap within the usual observational uncertainties (see Table~\ref{table:pop_mass}). The potential of this technique has only recently started to be exploited \citep{sd10,pdm10}.

For more massive models, we show how this type of diagram can help us identifying different convective core sizes, evolutionary stages, and therefore mixing prescriptions. Keeping in mind the results presented in Fig.~\ref{rat_large}, for a given mass value the diagram plotted in Fig.~\ref{popMass} can disentangle between different convective core sizes, as tracks of a given mass and composition neatly separate according to the mixing prescription applied. With an accurate measurement of the large frequency separation, as expected from the space missions, we can identify the size of the convective core and, as the mass and composition are known, the evolutionary stage.

Nevertheless, mass values determined from a combination of asteroseismic measurements and modelling are usually precise within 5-10\% when metallicity is known \citep{tm10,ng10}. In Table~\ref{table:pop_mass} we present the parameters of models for each mixing prescription with a mass difference of up to $\sim$8\%. These are neatly distinguishable in Fig.~\ref{popMass} for the considered error bars, showing that our method is able to break the degeneracy between mass value and mixing process. We stress again the fact that this analysis relies on known seismic and stellar parameters, in particular frequencies and metallicity. The effective temperature differences for the mass range considered in Table~\ref{table:pop_mass} within a mixing prescription are $\sim$150 K, a typical observational error bar. Stars with a larger mass difference for this composition will have an effective temperature not compatible with the observations at the same large frequency separation, which allows us to exclude them from the analysis.

As a closing remark, we notice that in Fig.~\ref{popMass} the track representing the evolution according to the Schwarzschild criterion shows abrupt jumps in its slope and mean values. This effect is a consequence of the convective process not properly taken into account at the edge of the core, and the abrupt variations correspond to the loops in the convective core profiles (see Sect.~\ref{ss:criteria}). However, the same general behavior as in the other mixing prescription is present, so a mean value can be considered for comparisons.
\section{Discussion}\label{s:disc}
While investigating different convective zone boundary definitions, we have seen that using the Schwarzschild criterion for models with growing convective cores leads to inconsistencies in the exact location of these regions. The layers at the top of the core may suddenly be mixed into the central region increasing the central hydrogen content and developing loops in the convective core boundary evolution (\textit{breathing pulses} phenomenon). The question that arises then is how to treat these zones. In the study made by \citet{cm82}, the authors took into account these layers by mixing them until convective neutrality according to the Schwarzschild criterion was recovered. The convective core grows beyond its previously established boundary and a new supply of fresh hydrogen is added to the core, also increasing the timespan a star spends transforming hydrogen into helium.

Another option for dealing with this \textit{breathing pulses} is using convective overshoot as implemented in GARSTEC, producing a smooth convective core profile during the main sequence evolution. This prescription does not modify the thermal structure of the overshooting layers. When an adiabatic stratification of the convective zone is considered instead, small fluctuations at the edge of the core still appear but of smaller size than the ones presented in Fig.~\ref{Ccore1} (M. Salaris, private communication). These fluctuations are not likely to affect the computation of adiabatic frequencies.

Overshooting is an interesting topic by itself, and a detailed analysis of it goes beyond the scope of this paper. However, we can mention two important aspects we think deserve attention. One is the temperature gradient in the overshooting region. We have assumed in our study that the temperature stratification in the radiative zone is not affected by the convective motions beyond the formal boundary, while convective penetration assumes an adiabatic stratification in this layer. Currently there is no consensus on the issue, and it is expected that numerical simulations and asteroseismology will cast some light on the topic \citep{mg07}. The other important aspect is the size of the overshooting region. An exponential decay of the convective velocities is an alternative to an adiabatic extension to a fraction of a scale height of the convective region \citep[see][for details]{pv07}. However, numerical simulations are not yet able to reproduce the conditions of stellar interiors \citep{bd09}.

Regarding semiconvection, our calibrated value for the the overshooting efficiency washes out the molecular weight discontinuity that produces semiconvective regions. Nevertheless, it has been suggested that the presence of a molecular weight barrier at the edge of the core might reduce the amount of overshooting from it \citep{vc99}. It might be possible for a semiconvective region to exist even with the presence of overshooting.

We have shown that the frequencies in the \textit{linear range} are sensitive to the central regions of the star. However, it is not straightforward to determine if the modes are reaching the convective core or not. The radial ($l=0$) modes propagate down to the center of the star, but the situation is different for the $l=1$ modes. Different estimations of the turning points can be made if the perturbation in the gravitational potential is neglected \citep[the cowling approximation,][]{tc41} or taken into account \citep{jj28}. In fact, it has been suggested that for stars with convective cores the dipolar modes reach the very center of the star \citep{mt05,mt06}. This is an interesting topic where further development is surely needed.

From the results presented in Sect.~\ref{s:ast} we can say that it is possible to estimate the size of the homogeneously mixed zone using seismic data. In particular, we can discriminate between stars with and without a convective core. There are a number of assumptions behind this prediction. Besides the expected frequency data, we assume that there are determinations of stellar parameters from either spectroscopy or photometry, with an uncertainty in effective temperature of $\sim$150 K. Given these observables we are only limited by the value of the stellar mass, which might be estimated within 5-10\% precision for a given mixing prescription using isochrones and asteroseismic scaling relations \citep{ds09a,sb10,ng10}.

The frequency ratios seem to sufficiently isolate the signal arising from the stellar core allowing us to study the central conditions of a star. Another asteroseismic tool aimed to study small convective cores was proposed by \citet{cm07}, where a combination between small and large separations was constructed and analyzed using modes of angular degree $l=0,1,2,3$. In their work, \citet{cm07} found that the sharp discontinuity formed at the edge of the convective core has a direct impact on the slope of their proposed quantity when plotted against frequency, showing that the absolute value of the slope increases as a function of the size of the chemical discontinuity (and therefore with decreasing amount of central hydrogen content). Since observing $l=3$ modes is very challenging due to severe amplitude reduction, \citet{cb10} extended this study using the small frequency separations defined in Eqs.~\ref{eqn:d01} and \ref{eqn:d10} and analyzing the changes in the slope, reaching the same conclusions as \citet{cm07}.

The behavior with evolution is the same for the small frequency separations, the frequency ratios, and the asteroseismic tool proposed by \citet{cm07}. Thus, our findings are applicable to all of them. We have shown in Sect.~\ref{ss:rat_ms} that, although there is a relation between the size of the discontinuity in the adiabatic sound speed and the absolute slope of the ratios, the total extent of the convective core and the central hydrogen content heavily influence the frequency ratios. Three models with the same large frequency separation and similar size of chemical discontinuity at the edge of the core have very different absolute slopes due to their different convective core sizes and central hydrogen contents. On the other hand, models which barely change the size of the convective core during the hydrogen burning phase show the mildest change in the absolute slope of the ratios.

From the analysis performed in this paper, we have a clear theoretical understanding of the different contributions in the \textit{linear range} of the frequency domain and how to disentangle between the effects of mass and type of mixing for realistic error expectations in the observables. We have based our study of the frequency ratios assuming that the oscillations are pure p-modes, but this assumption breaks down when modes with mixed p-mode and g-mode character appear \citep[see][and references therein]{ca10}. These mixed modes carry information from the deep interior and are very sensitive to the core conditions, making them ideal probes for age calibrations in sub-giant stars and mixing processes occurring during the main sequence \citep[e.g][]{dp91,cs05,dm10,tm10}. For instance, they could be detected in main sequence F-type stars with shallow convective envelopes, where this avoided crossings could more easily reach the visible frequency range. However, different analysis tools from the ones presented in this paper must be developed to fully exploit their potential.
\section{Summary and conclusions}\label{s:conc}
We have presented results of our study of the influence of different convective boundary definitions and prescriptions of chemical mixing on main sequence evolution of low-mass stars. This has important consequences for the overall evolution of low-mass stars. We have focused our study on the effects on stellar ages and CMD morphology of the existence and size of convective cores.

The age of a star at the end of the main sequence phase depends on the amount of fuel available for core burning. In the prescriptions we have considered, this is governed by the size of the homogeneously mixed zone in the center. The implementation of the Ledoux criterion for convection and no mixing beyond the formal boundaries of the central convective region produces the smallest size of convective core and consequently the shortest main sequence age for a given stellar mass and composition. The time-dependent diffusive mixing in the semiconvective region increases the amount of hydrogen in the center, reaching the same convective core size as the Schwarzschild criterion for high enough mixing efficiencies. Convective overshoot can increase even further the extension of the mixed zone and the associated age value at the end of the main sequence.

The appearance of a hook-like feature at the end of the central hydrogen burning phase can be shifted to higher mass values by the inclusion of the Ledoux criterion, as we have shown for a 1.2 M$_{\sun}$ model. This has relevant consequences when isochrones are used to reproduce the shape of the CMD of a stellar population given a chemical composition and a set of input physics \citep{dv07,zm10}. Previous methods of analysis do not allow to ensure the existence or absence of a semiconvective region in stellar interiors. Further development of numerical simulations and asteroseismic tools, such as the one presented in this paper, can be the key to solve this interesting issue \citep{gb07,am08}. In the same line, constraining the existence and size of the central mixed region is another step towards a comprehensive understanding of mixing processes. 

We have focused our attention upon using frequency combinations sensitive to the convective core size and the evolutionary stage of the star to discriminate between the different mixing prescriptions. For this purpose, we have chosen the frequency ratios and analyzed their behavior throughout the main sequence evolution for different masses and mixing prescriptions. The ratios present a low-frequency domain which is governed by the global seismic properties of the star (the large frequency separation). However, for higher frequencies the behavior of the ratios changes together with the inner conditions of the star. These are all high order low-degree modes and correspond to what we have termed the \textit{linear range}.

Using this \textit{linear range}, our findings predict that we can discriminate between stars with and without convective cores, constrain the size of the homogeneously mixed central region and disentangle the effects from stellar mass. All of these provided we use an adequate frequency combination to isolate the signal coming from the core. The method relies on our knowledge of the stellar composition, where large uncertainties can still be present. In order to fully assess this and other potentially important effects, such as helium abundance and microscopic diffusion, a hare and hounds exercise with extension to other metallicities and applications to real data is currently underway to further test the method (Silva Aguirre et al., in preparation). Nevertheless, the extensive and highly accurate sets of observations obtained by the CoRoT and \textit{Kepler} missions \citep[e.g][]{em08,ds10,wc10,jcd10b} open an exciting possibility for stellar physics: to apply differential analysis to stars with a common property, such as metallicity, and pin down the differences arising from the inner layers of their structure.
\begin{acknowledgements}
The authors would like to the anonymous referee for useful suggestions that helped improving the manuscript. They would also like to thank S. Charpinet and M. Salaris for providing evolutionary models and pulsation frequencies for comparisons.
\end{acknowledgements}
\bibliographystyle{aa} 
\bibliography{references} %
\end{document}